\documentclass[]{mn2e}
\usepackage{epsfig}
\usepackage{graphicx}
\usepackage{float}

\title[SFH in the MCs using LPVs]{The star formation history of the Magellanic
Clouds derived from long-period variable star counts}
\author[Rezaei kh. et al.]{
Sara Rezaei kh.$^1$,
Atefeh Javadi$^1$,
Habib Khosroshahi$^1$ and
Jacco Th.\ van Loon$^2$\\
$^1$School of Astronomy, Institute for Research in Fundamental Sciences (IPM),
    P.O.\ Box 19395-5531, Tehran, Iran\\
$^2$Lennard-Jones Laboratories, Keele University, ST5 5BG, UK}
\pagerange{\pageref{firstpage}--\pageref{lastpage}}
\pubyear{2014}
\begin{document}
\maketitle
\label{firstpage}
\begin{abstract}
We present the first reconstruction of the star formation history (SFH) of the
Large and Small Magellanic Clouds (LMC and SMC) using Long Period Variable
stars. These cool evolved stars reach their peak luminosity in the
near-infrared; thus, their K-band magnitudes can be used to  derive their
birth mass and age, and hence the SFH can be obtained. In the LMC, we found a
10-Gyr old single star formation epoch at a rate of $\sim1.5$ M$_\odot$
yr$^{-1}$, followed by a relatively continuous SFR of $\sim0.2$ M$_\odot$
yr$^{-1}$, globally. In the core of the LMC (LMC bar), a secondary, distinct
episode is seen, starting 3 Gyr ago and lasting until $\sim0.5$ Gyr ago. In
the SMC, two formation epochs are seen, one $\sim6$ Gyr ago at a rate of
$\sim0.28$ M$_\odot$ yr$^{-1}$ and another only $\sim0.7$ Gyr ago at a rate of
$\sim0.3$ M$_\odot$ yr$^{-1}$. The latter is also discernible in the LMC and
may thus be linked to the interaction between the Magellanic Clouds and/or
Milky Way, while the formation of the LMC bar may have been an unrelated
event. Star formation activity is concentrated in the central parts of the
Magellanic Clouds now, and possibly has always been if stellar migration due
to dynamical relaxation has been effective. The different initial formation
epochs suggest that the LMC and SMC did not form as a pair, but at least the
SMC formed in isolation.
\end{abstract}

\section{Introduction}

The Magellanic Clouds (Large, LMC, and Small, SMC) are irregular, gas-rich
dwarf galaxies in our Local Group. The presence of a bar-like structure --
especially prominent in the LMC -- suggests that they could have originally
been barred spiral galaxies transformed into present-day irregular galaxies
due to their tidal interactions with one another and with the Milky Way. At a
distance of 48.5 kpc (Macri et al.\ 2006; Freedman et al.\ 2010), the LMC is
the third closest galaxy to the Milky Way. With an inclination of $32^\circ$
(Haschke et al.\ 2011) it offers us an excellent view of its structure and
content. The SMC is located at a distance of $\sim61$ kpc (Hilditch et al.\
2005) with an inclination of only $2.6^\circ$ (Subramanian et al.\ 2011).

Probing the star formation history (SFH) in galaxies informs us about galaxy
formation and evolution. Star clusters revealed differences in the SFH between
the LMC and the Milky Way (Hodge 1960; Sagar \& Pandey 1989): the LMC contains
a larger fraction of young clusters relative to the Milky Way. Other methods
of determining the SFH include a comparison of the observed Colour--Magnitude
Diagrams (CMDs) with the model-built CMDs (e.g., Bertelli et al.\ 1992).
Harris \& Zaritsky (2009) used multi-color photometry of millions of stars
over $\approx65$ deg$^2$ of the LMC to derive the SFH; they found multiple
episodes of star formation between 5 to 0.1 Gyr ago in addition to an initial
burst of star formation $\sim12$ Gyr ago. In a recent study, Cignoni et al.\
(2013) studied various fields within the bar and wing of the SMC and confirmed
a dominant intermediate-age star formation epoch. Their derived metallicity
for the fields support a well mixed metal content of the SMC. The CMD approach
has also revealed that the most recent star formation, as young as
$\approx100$ Myr, is confined to the central regions in both the LMC and SMC
(Indu \& Subramanian 2011).

Our approach to investigate the SFH is based on employing Long Period Variable
stars (LPVs). These are mostly Asymptotic Giant Branch (AGB) stars at their
very late stage of evolution, as well as more massive red supergiants (RSGs).
The AGB stars alone have low- to intermediate-mass ranges which relate to
diverse ages and hence time epochs in the past, and they can thus be used to
derive Star Formation Rates (SFRs) over much of a galaxy's cosmological
evolution. The AGB stars develop unstable mantles which cause radial pulsation
at periods of one or a few years; they are luminous ($\sim10^4$ L$_\odot$) and
cool ($<4000$ K) and hence stand out at near-infrared wavelengths. Their
colours can be further reddened by enhanced atmospheric opacity due to
carbon-dominated chemistry (in carbon stars) or due to attenuation by
circumstellar dust produced in their winds. Thanks to recent large-scale,
extended monitoring projects such as OGLE-I, II, III, MACHO and others, we are
now in a position to use LPVs to study these nearby galaxies.

The data and the methodology are described in Section 2. In Section 3 we
present our analysis and the results. In Section 4 we present the SFH derived
from our analysis. Section 5 presents a discussion of the findings and a
comparison with other studies. Section 6 contains a summary and conclusions.

\section{Data and Methodology}

In this paper, we use the catalogue of LPVs in the LMC from Spano et al.\
(2011) and in the SMC from Soszy\'nski et al.\ (2011). In the case of the LMC,
they used the EROS-2 survey which started functioning efficiently from 1996;
it employed two cameras enabling observations at two wavelengths, blue (BE)
and red (RE), simultaneously. After seven years of monitoring, it provided a
database of 856,864 variables in the LMC of which 43,551 are considered LPVs
(Spano et al.\ 2011). The EROS-2 survey covered an area of 88 deg$^{2}$ on the
LMC, which is a superior coverage in comparison with other projects such as
MACHO and OGLE-III which covered areas on the LMC of 13.5 deg$^{2}$ and 40
deg$^{2}$, respectively.

Spano et al.\ cross-identified the EROS-2 LPV candidates with variable stars
from the OGLE-III and MACHO surveys. They found nearly 60\% of LPVs in common
between EROS-2 and MACHO; the difference is attributed to differences in sky
coverage between the surveys (Spano et al.\ 2011). For the SMC, Soszy\'nski et
al.\ used the OGLE-III project which has monitored an area of 14 deg$^{2}$ on
the SMC. After 13 years of monitoring in the optical V- and near-infrared
I-band, the OGLE survey has provided a database of about 6 million stars in
the SMC of which 19,384 are identified as LPVs are detected (Soszy\'nski et
al.\ 2011).

As mentioned, AGB stars become surrounded by dust, which causes reddening and
makes them particularly faint in optical bands. Therefore, infrared (IR)
studies are preferred. For a comparison and more detailed study of the core
region we also utilise Ita et al.'s (2004a) data of variable stars in the
Magellanic Clouds. This catalogue had been obtained by cross-identifying the
OGLE-II and SIRIUS data (simultaneously observated in J-, H-, and K-band). The
total area covered by the SIRIUS survey is 3 deg$^{2}$ and 1 deg$^{2}$ in the
LMC and SMC, respectively, covering the core region of these galaxies -- in
particular, the bar of the LMC (see Fig.\ 1). Spano et al.\ (2011)
cross-matched the EROS-2 survey with 2MASS near-IR (and {\it Spitzer} mid-IR)
data, within a search radius of $1^{\prime\prime}$. For the SMC, we cross-matched
the original catalogue of Soszy\'nski et al.\ (2011) with 2MASS, also within a
search radius of $1^{\prime\prime}$.

\begin{figure*}
\centerline{\hbox{
\psfig{figure=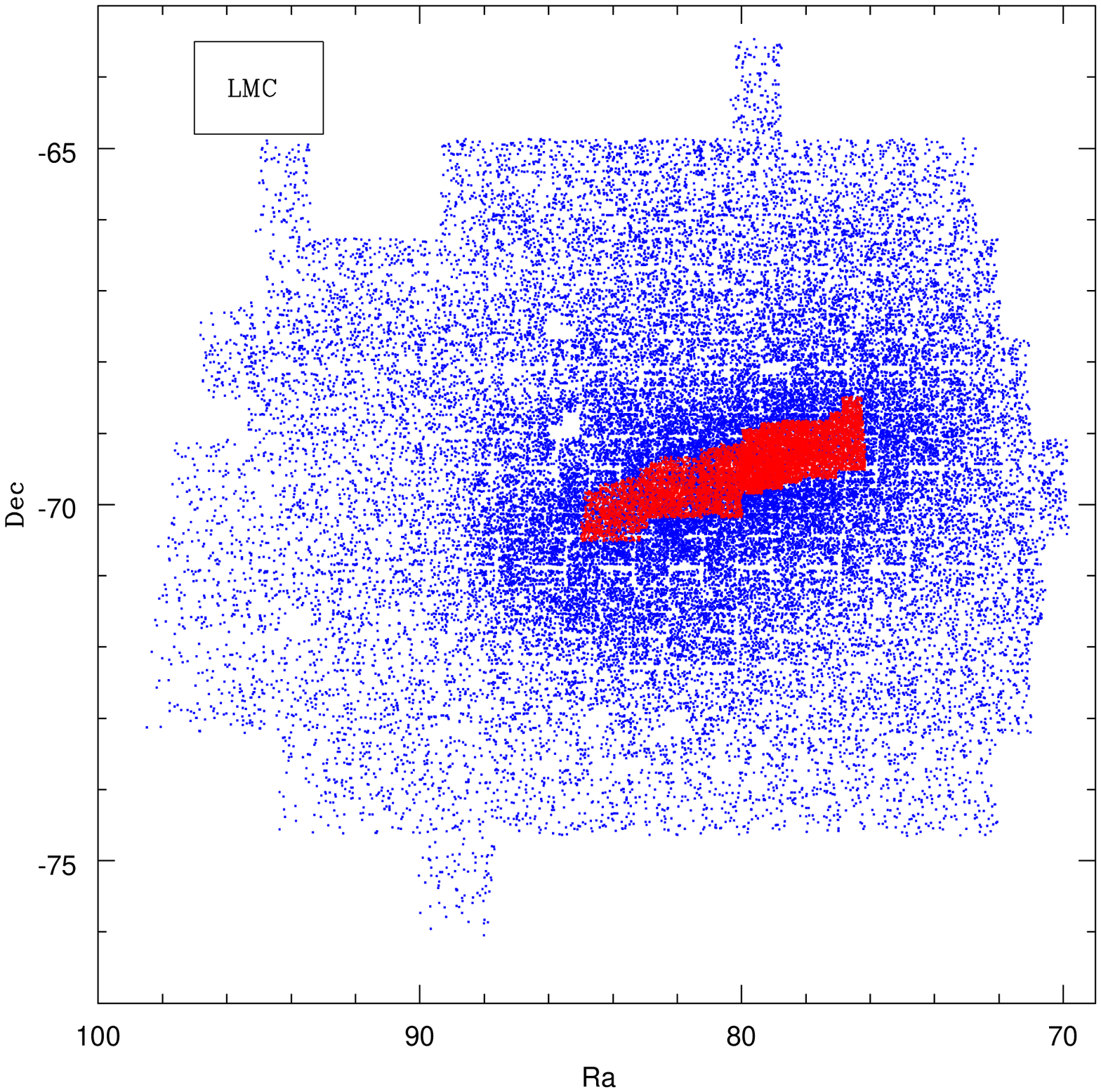,width=88mm}
\psfig{figure=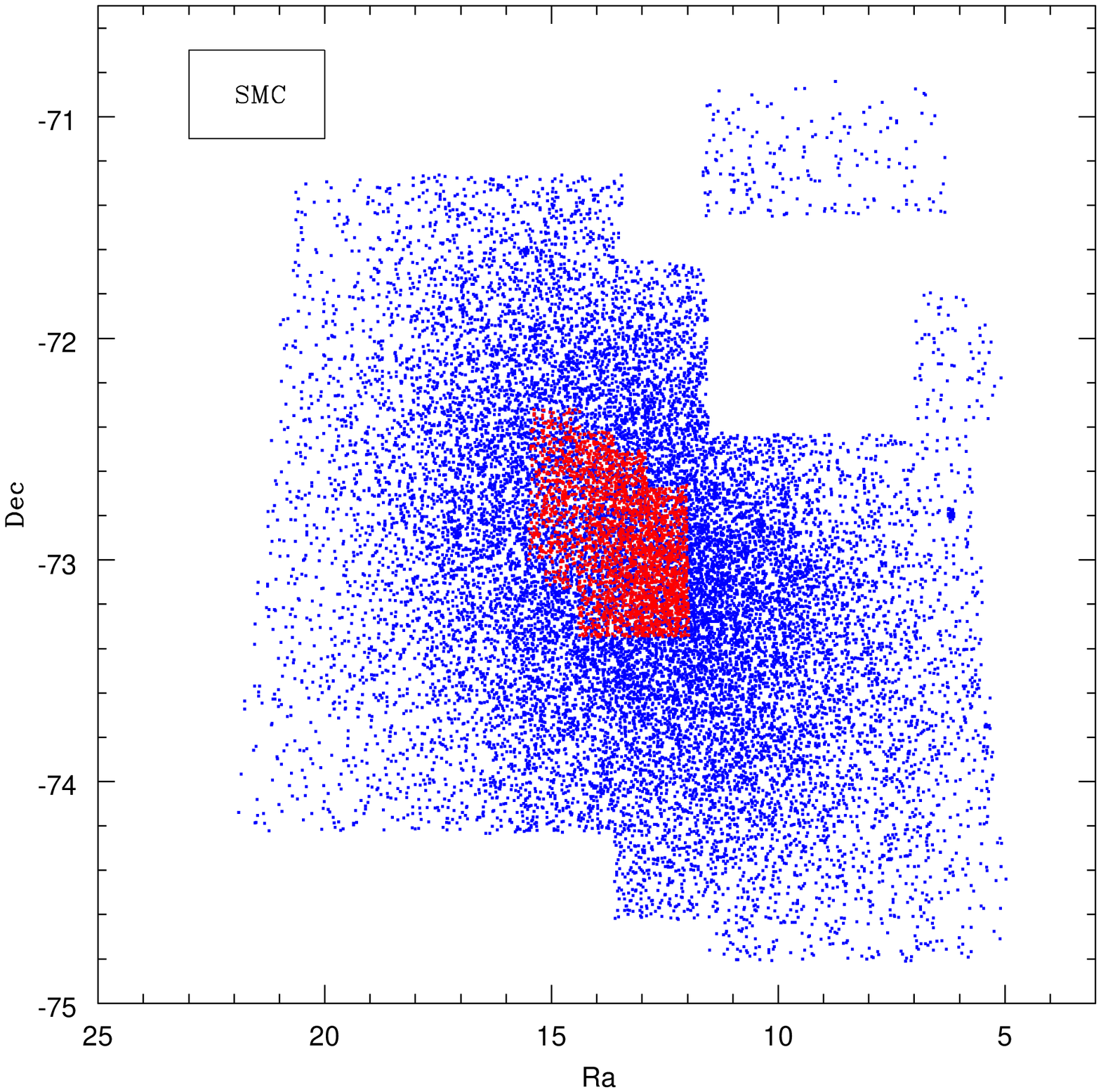,width=88mm}
}}
\caption[]{{\it Left:} spatial coverage of the Spano et al.\ (2011) survey
(blue) and Ita et al.\ (2004a) survey (red) in the LMC. {\it Right:} spatial
coverage of the Soszy\'nski et al.\ (2011) survey (blue) and Ita et al.\
(2004a) survey (red) in the SMC.}
\end{figure*}

Ita et al.\ classified the variable stars into nine groups for the LMC and
eight groups for the SMC, based on their location in the period--K-band
magnitude plane. The categories include Cepheids, Miras, semi regulars,
irregular variables and eclipsing binaries (Ita et al.\ 2004a). In total, the
number of variable stars in Ita's catalogues of the LMC and SMC are 8852 and
2927, respectively. For the purpose of this study, we select LPVs pulsating in
the fundamental mode, equal to sequences C and D in Ita et al.\ (2004a) above
the (first ascent) Red Giant Branch (RGB) tip ($K\sim12$ and $\sim12.7$ mag
for the LMC and SMC, respectively). Spano et al.\ (2011) classified their
variable stars using period--luminosity relations ($\log P$--$K$ diagram),
into six groups (Figs.\ 22 and 23 in their paper), from which sequence C is
considered as Miras pulsating in the fundamental mode. Sequence D comprises
mostly of stars with secondary long periods that are probably not due to the
same radial pulsation we are after, but it does contain some dusty LPVs that
have lengthened pulsation periods as a result of diminished mass (Ita et al.\
2004b; Spano et al.\ 2011). For the purpose of this study we select variables
belonging to sequence C, and variables with "only" one period in sequence D.

\begin{figure*}
\centerline{\hbox{
\psfig{figure=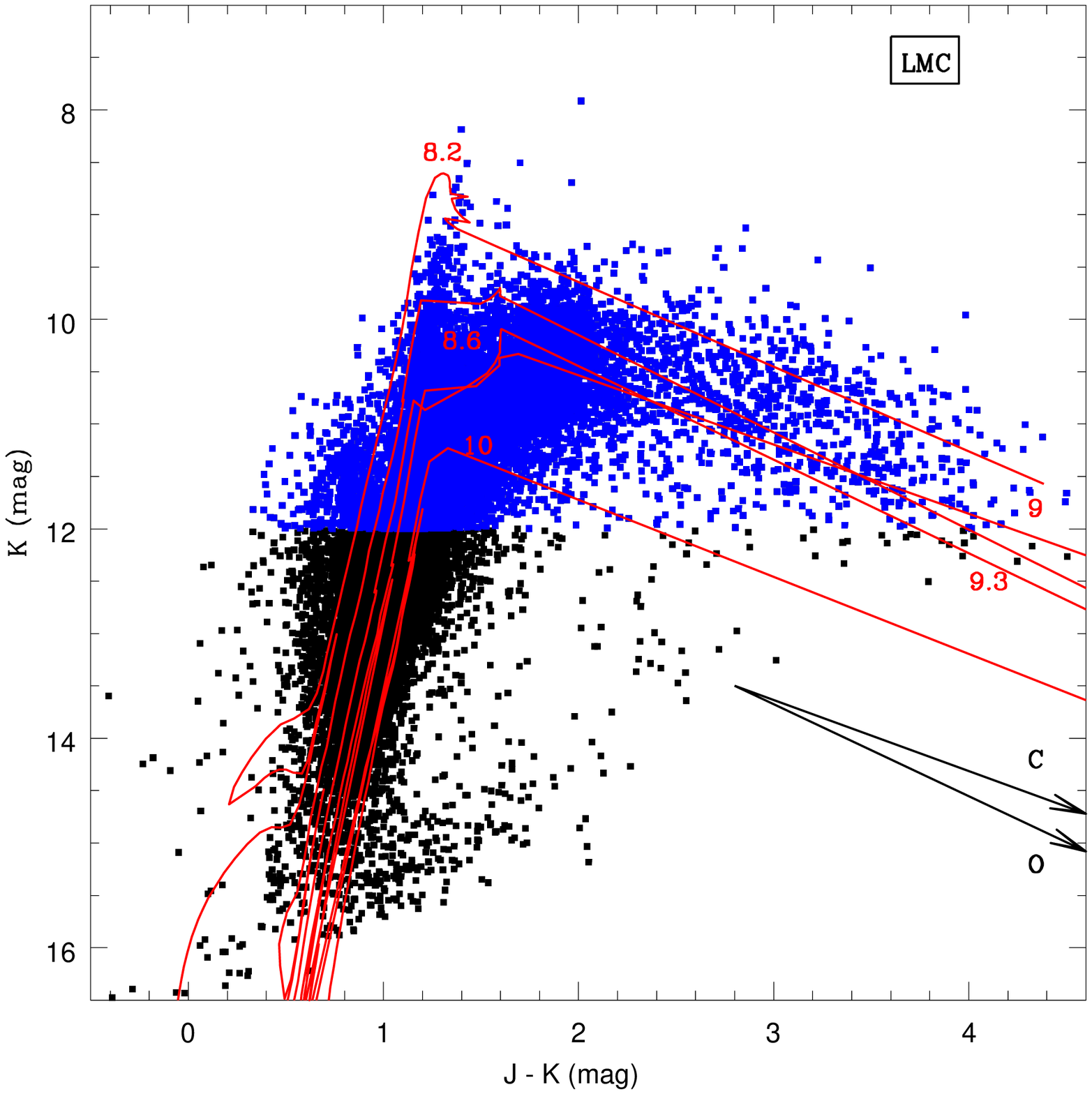,width=88mm}
\psfig{figure=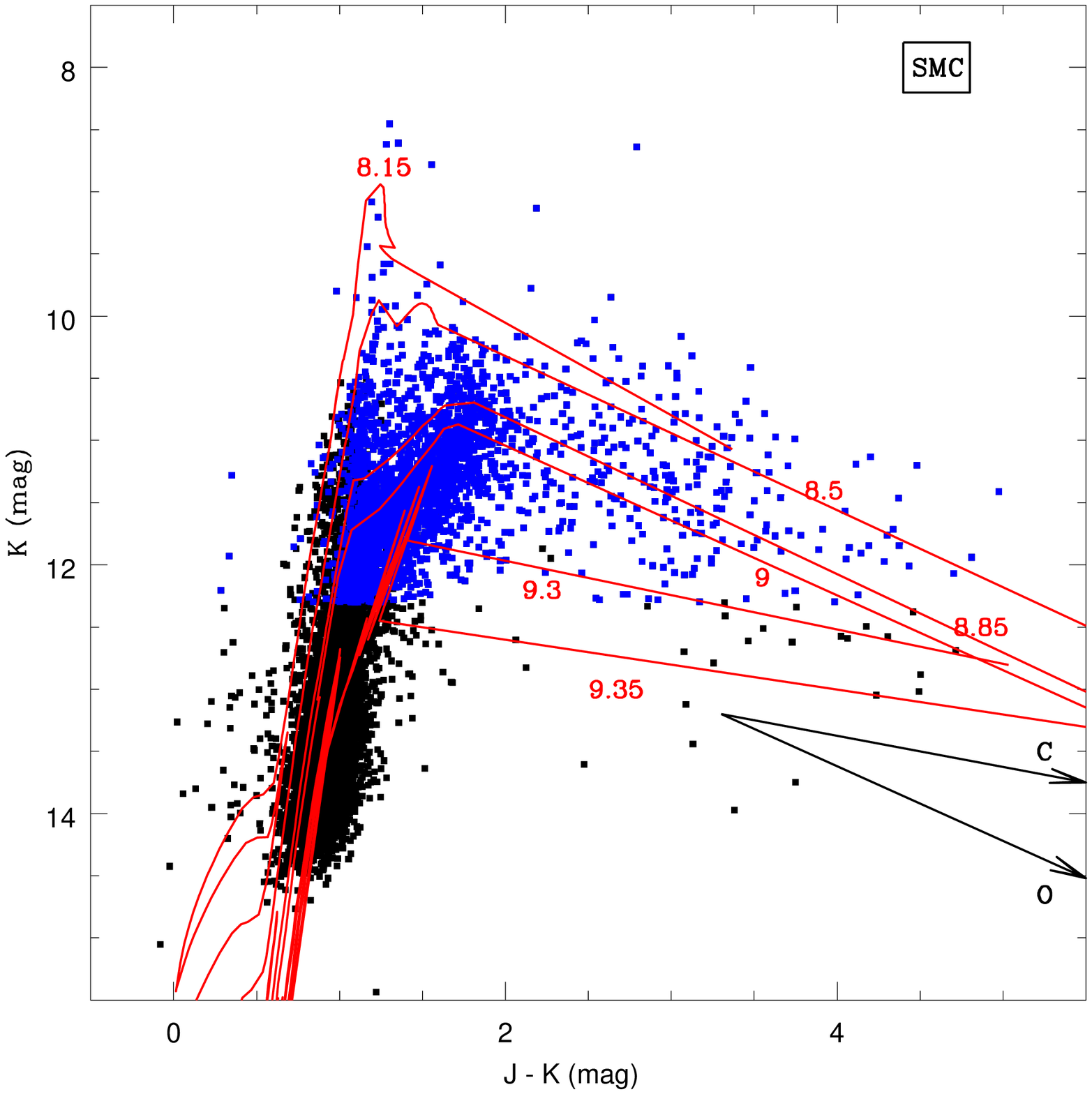,width=88mm}
}}
\caption[]{CMD of the LMC ({\it left}) and SMC ({\it right}), overlain with
Marigo et al.\ (2008) isochrones (labelled by logarithmic age in yr) ; among with circumstellar reddening vectors for oxygenous and carbonaceous dust (O and C, respectively), explained in Section 3.1.1. Blue points are stars above the RGB-tip.}
\end{figure*}

An appropriate stellar evolution model is provided by the Padova group (Marigo
et al.\ 2008). LPVs are expected to be at the end-points of the AGB (or RSG)
part of the isochrones in CMDs. Thus, the Padova isochrones were fitted to the
CMDs of the Magellanic Clouds in order to correct the reddened LPVs (see Fig.\
2). Two different groups of reddening slopes are seen, those with steeper
slopes are associated with those stars surrounded by oxygeneous dust whilst
some with shallower slopes are associated with carbonaceous dust. Not all
stars are reddened; the correction is applied only to those stars which have
$J-K>1.5$ mag.

\section{From K-band magnitude to the star formation history}

The total mass of stars formed between $t$ and $t+{\rm d}t$
depends on star formation rate, $\xi(t)$, as
\begin{equation}
{\rm d}M(t)\ =\ \xi(t)\,{\rm d}t.
\end{equation}
Thus, $N$, the number of stars formed, is defined by the following equation:
\begin{equation}
{\rm d}N(t)\ =\ \left[\frac
{\int_{\rm min}^{\rm max}f_{\rm IMF}(m)\,{\rm d}m}
{\int_{\rm min}^{\rm max}f_{\rm IMF}(m)m\,{\rm d}m}
\right]\,{\rm d}M(t),
\end{equation}
in which $f_{\rm IMF}$ is the initial mass function. The nominator represents
the total number of stars within the mass range of $M_{\rm min}$--$M_{\rm max}$,
whilst the denominator gives their total mass. Therefore, the fraction refers
to the number of stars per mass unit. For the initial mass function, the
Kroupa (2001) model was used where we assumed $M_{\rm min}=0.02$ M$_\odot$ and
$M_{\rm max}=200$ M$_\odot$. The number of stars formed between $t$ and
$t+{\rm d}t$ is then obtained as follows:
\begin{equation}
{\rm d}n(t)\ =\ \left[\frac
{\int_{m(t)}^{m(t+dt)}f_{\rm IMF}(m)\,{\rm d}m}
{\int_{\rm min}^{\rm max}f_{\rm IMF}(m)\,{\rm d}m}
\right]\,{\rm d}N(t).
\end{equation}

Because the large-amplitude, long-period variability only comprises a short
phase in the evolution of a star lasting $\delta t$, the number of such
variable stars that were formed is reduced to
\begin{equation}
{\rm d}n^\prime(t)\ =\ \left[\frac{\delta t}{{\rm d}t}\right]\,{\rm d}n(t).
\end{equation}

Thus, the final equation for the star formation rate derived from the variable
star counts is:
\begin{equation}
\xi(t)\ =\  \left[\frac
{\int_{\rm min}^{\rm max}f_{\rm IMF}(m)m\,{\rm d}m}
{\int_{m(t)}^{m(t+{\rm d}t)}f_{\rm IMF}(m)\,{\rm d}m}
\right]\,\frac{{\rm d}n^\prime(t)}{\delta t}.
\end{equation}

\subsection{From K-band magnitude to age}

For reasons that will become clear below, we convert K-band magnitudes into
stellar birth mass, first, before converting that into the age of the star.
For the latter, we use the following mass--age relation:
\begin{equation}
\log{t\,{\rm [yr]}}\ =\ a\,\log{M\,{\rm [M}_\odot{\rm]}}+b,
\end{equation}
where $a$ and $b$ are given in Javadi et al.\ (2011), based on the models from
Marigo et al.\ (2008).

Javadi et al.\ (2011) give the mass--luminosity (K-band magnitude) relation as
\begin{equation}
\log{M\,{\rm [M}_\odot{\rm]}}\ =\ aK+b,
\end{equation}
where $K$ is the (extinction-corrected) K-band magnitude and $a$ and $b$ are
defined in Table 1 for various ranges of $K$. We use a distance modulus of
$\mu=18.5$ mag and $\mu=18.9$ mag for the LMC and SMC, respectively.

\begin{table}
\caption[]{Mass--luminosity relation, $\log{M\,{\rm [M}_\odot{\rm ]}}=aK+b$,
valid for the Large ($Z=0.008$) and Small ($Z=0.004$) Magellanic Clouds
(Javadi et al.\ 2011).}
\begin{tabular}{ccc}
\hline\hline
\multicolumn{3}{c}{$Z=0.008$} \\
\hline
$a$              & $b$                     & validity range       \\
\hline
$-0.840$         &  \llap{1}2.840          &        $K\leq7.332$  \\
$-0.589$         &          9.391          &  $7.332<K\leq8.387$  \\
$-0.188\pm0.028$ &         $3.465\pm0.043$ &  $8.387<K\leq8.900$  \\
$-0.142\pm0.027$ &         $2.758\pm0.041$ &  $8.900<K\leq9.413$  \\
$-0.188\pm0.022$ &         $3.487\pm0.036$ &  $9.413<K\leq9.927$  \\
$-0.501\pm0.016$ &         $8.183\pm0.024$ &  $9.927<K\leq10.440$ \\
$-0.248\pm0.018$ &         $4.335\pm0.026$ & $10.440<K\leq10.953$ \\
$-0.128\pm0.025$ &         $2.257\pm0.035$ &        $K>10.953$    \\
\hline\hline
\multicolumn{3}{c}{$Z=0.004$} \\
\hline
$a$              & $b$                     & validity range       \\
\hline
$-0.708$         &  \llap{1}1.1268         &        $K\leq8.734$  \\
$-0.209\pm0.051$ &         $3.783\pm0.077$ &  $8.734<K\leq9.140$  \\
$-0.240\pm0.054$ &         $4.244\pm0.078$ &  $9.140<K\leq9.545$  \\
$-0.050\pm0.060$ &         $1.291\pm0.091$ &  $9.545<K\leq9.951$  \\
$-0.131\pm0.054$ &         $2.583\pm0.083$ &  $9.951<K\leq10.356$ \\
$-0.243\pm0.041$ &         $4.409\pm0.057$ & $10.356<K\leq10.762$ \\
$-0.714\pm0.039$ & \llap{1}$2.310\pm0.055$ & $10.762<K\leq11.167$ \\
$-0.109\pm0.043$ &         $1.932\pm0.070$ & $11.167<K\leq11.573$ \\
$-0.153\pm0.046$ &         $2.690\pm0.067$ &        $K>11.573$    \\
\hline
\end{tabular}
\end{table}

\subsubsection{Dust correction}

The light from the variable stars can be attenuated by interstellar and/or
circumstellar dust; this is wavelength dependent resulting also in reddening
of the near-IR colours. To determine the intrinsic K-band magnitude, the
photometry needs to be de-reddened. The (de-)reddening slope in the CMD
depends on whether the dust is oxygenous or carbonaceous in composition. Stars
with a birth mass in the range of 1.5--4 M$_\odot$ are expected to have become
carbon stars due to the third dredge-up of nculear-processed material material
with a carbon:oxygen ratio in excess of unity. In stars with a birth mass 
$<1.5$ M$_\odot$, third dredge-up is not sufficiently efficient, whilst, for
stars with a birth mass $>4$ M$_\odot$ nuclear burning of the carbon at the
bottom of the convection zone prevents the surface to be enriched in carbon.
The reddening correction equation is
\begin{equation}
K\ =\ K+a(1.25-(J-K)),
\end{equation}
where $a$ is the average slope for each dust type.

The procedure followed was to first apply a correction assuming carbonaceous
dust. If the mass resulting from this corrected K-band magnitude ended up in
the range 1.5--4 M$_\odot$ then the star was treated as a carbon star.
Otherwise, the correction was replaced by one assuming oxygenous dust, and a
mass was derived accordingly.

Figure 3 shows the mass histogram for different choices of reddening
correction. To appreciate the impact of these choices on the derived SFH, we
also applied the reverse approach; assuming that stars are oxygen-rich unless
their masses would fall within 1.5--4 M$_\odot$.

\begin{figure*}
\centerline{\hbox{
\psfig{figure=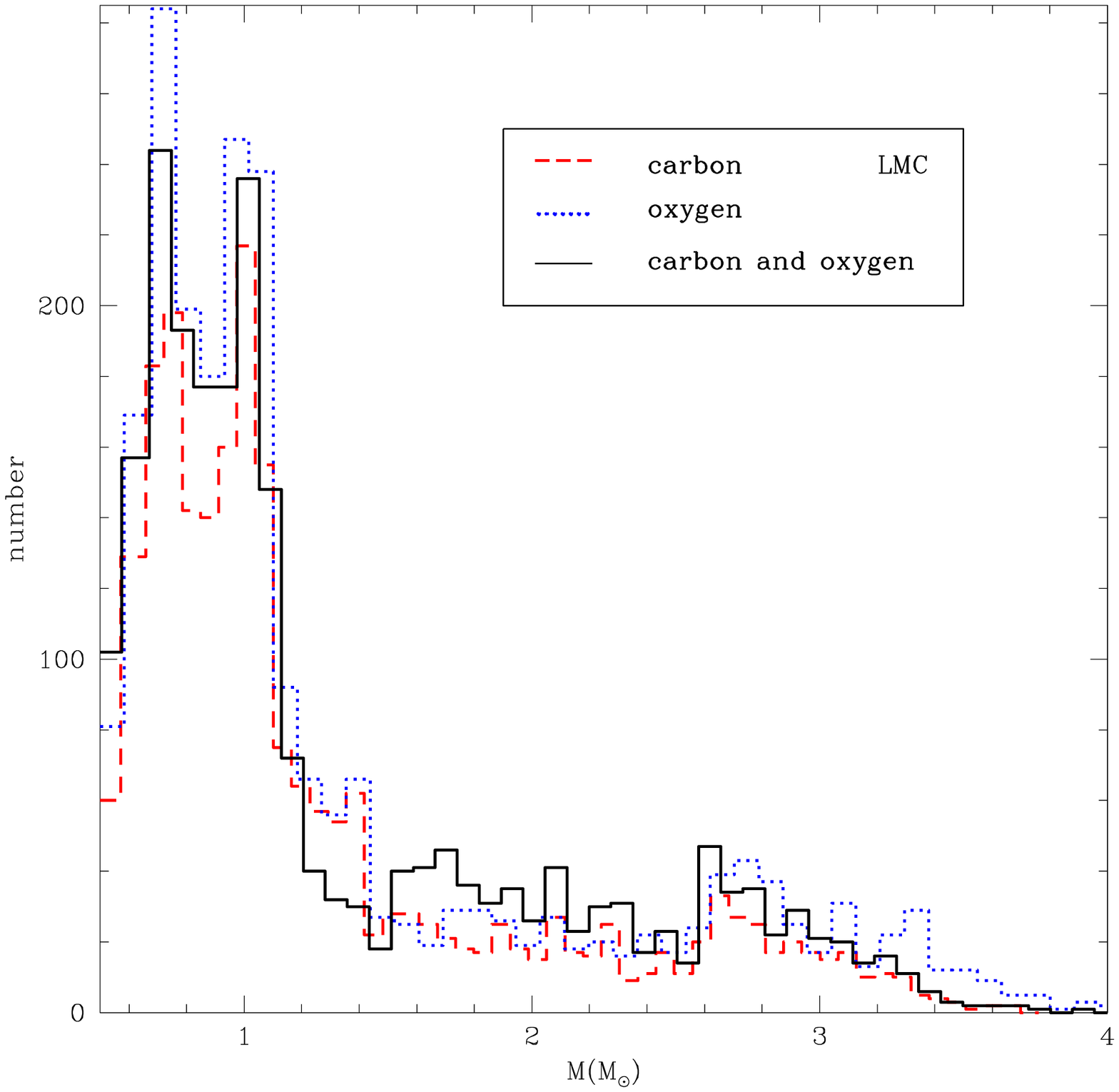,width=88mm}
\psfig{figure=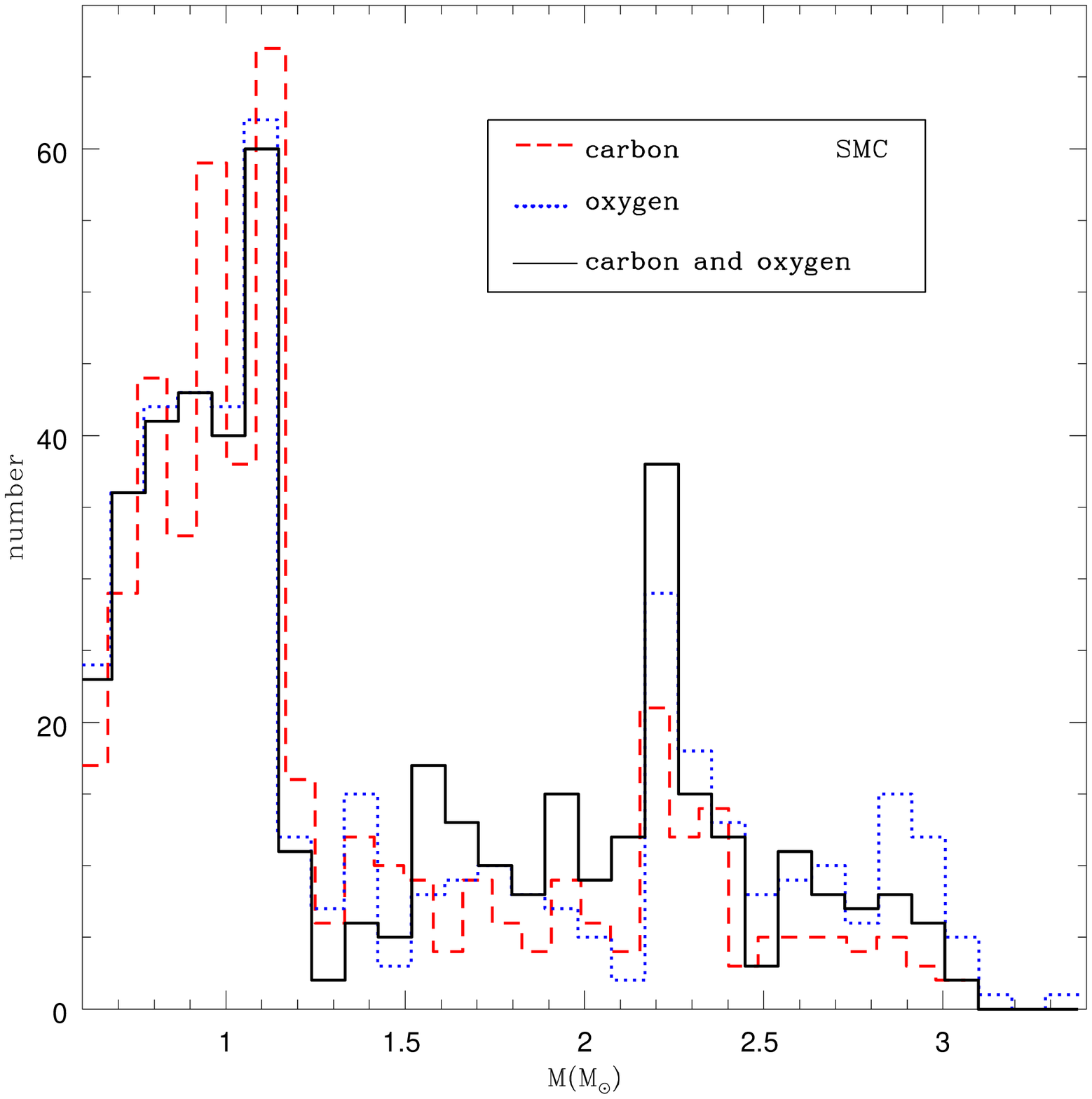,width=88mm}
}}
\caption[]{Present day mass function of the LMC ({\it left}) and SMC ({\it
right}). The red dashes are related to carbonaceous correction only, the blue
dots are for oxygenous correction only, and the black lines refer to the
approach described in the text applying carbonaceous and oxygenous corrections
depending on the resulting mass.}
\end{figure*}

\subsection{The fraction of the time spent pulsating}

Eq.\ (5) depends on a correction factor ($\delta t$) for the duration of the
evolutionary phase in which the star exhibits large-amplitude, long-period
pulsations. Generally, since a massive star evolves faster, its pulsation
duration is shorter. Thus, low mass stars are more likely to be identified as
variables, in comparison to more massive stars. However, the pulsation phase
depends sensitively on the star's effective temperature which, in turn,
depends on metallicity, mass loss, et cetera.

\begin{table}
\caption[]{Parameters adopted for the mass--pulsation relation,
$\log(\delta t/t) = D + \sum_{i=1}^{3}a_i
\exp\left( (\log{M\,{\rm [M}_\odot{\rm ]}}-b_i)^2/(2c_i^2)\right)$, where
$\delta t$ is the pulsation duration and $t$ is the age of the star, for the
Large ($Z=0.008$) and Small ($Z=0.004$) Magellanic Clouds (Javadi et al.\
2011).}
\begin{tabular}{ccccc}
\hline\hline
\multicolumn{5}{c}{$Z=0.008$} \\
\hline
$D$     & $i$ & $a$  & $b$   & $c$   \\
\hline
$-3.96$ &  1  & 2.34 & 1.281 & 0.378 \\
        &  2  & 1.32 & 0.460 & 0.165 \\
        &  3  & 0.38 & 0.145 & 0.067 \\
\hline\hline
\multicolumn{5}{c}{$Z=0.004$} \\
\hline
$D$     & $i$ & $a$  & $b$   & $c$   \\
\hline
$-4.00$ &  1  & 2.29 & 1.217 & 0.408 \\
        &  2  & 0.84 & 0.524 & 0.093 \\
        &  3  & 0.87 & 0.206 & 0.088 \\
\hline
\end{tabular}
\end{table}

The mass--pulsation relation from Javadi et al.\ ( 2011) is used to connect
the birth mass of stars to their pulsation duration. Relative to its age, it
is parameterised as
\begin{equation}
\log\left(\frac{\delta t}{t}\right)\ =\ D + \sum_{i=1}^{3}a_i \exp
\left[\frac{\left(\log{M\,{\rm [M}_\odot{\rm ]}}-b_i\right)^2}{2c_i^2}\right],
\end{equation}
where $a$, $b$, $c$ and $D$ are defined in Table 2.

Given the above argument, the star formation rate estimated using LPVs is
shown to be underestimated by a factor of 10 (Javadi et al.\ 2013) in
comparison to the star formation rate obtained using $H_\alpha$ emission, for
instance. This inaccuracy was exposed by a mismatch between the integrated
mass loss and birth mass. To reconcile both the mass loss budget and star
formation rate with independent evidence, the pulsation duration adopted from
the Marigo et al.\ (2008) models needs to be decreased by a factor 10. This is
a fairly robust result and we thus apply the correction in our work.  Given the uncertainties in the models, in particular with regard to the duration of the LPV phase, the star formation rates and stellar masses formed are possibly uncertain by up to a factor of a few.

\section{Results and discussion}
\subsection{The global Star Formation History of the Magellanic Clouds}

\begin{figure*}
\centerline{\hbox{
\psfig{figure=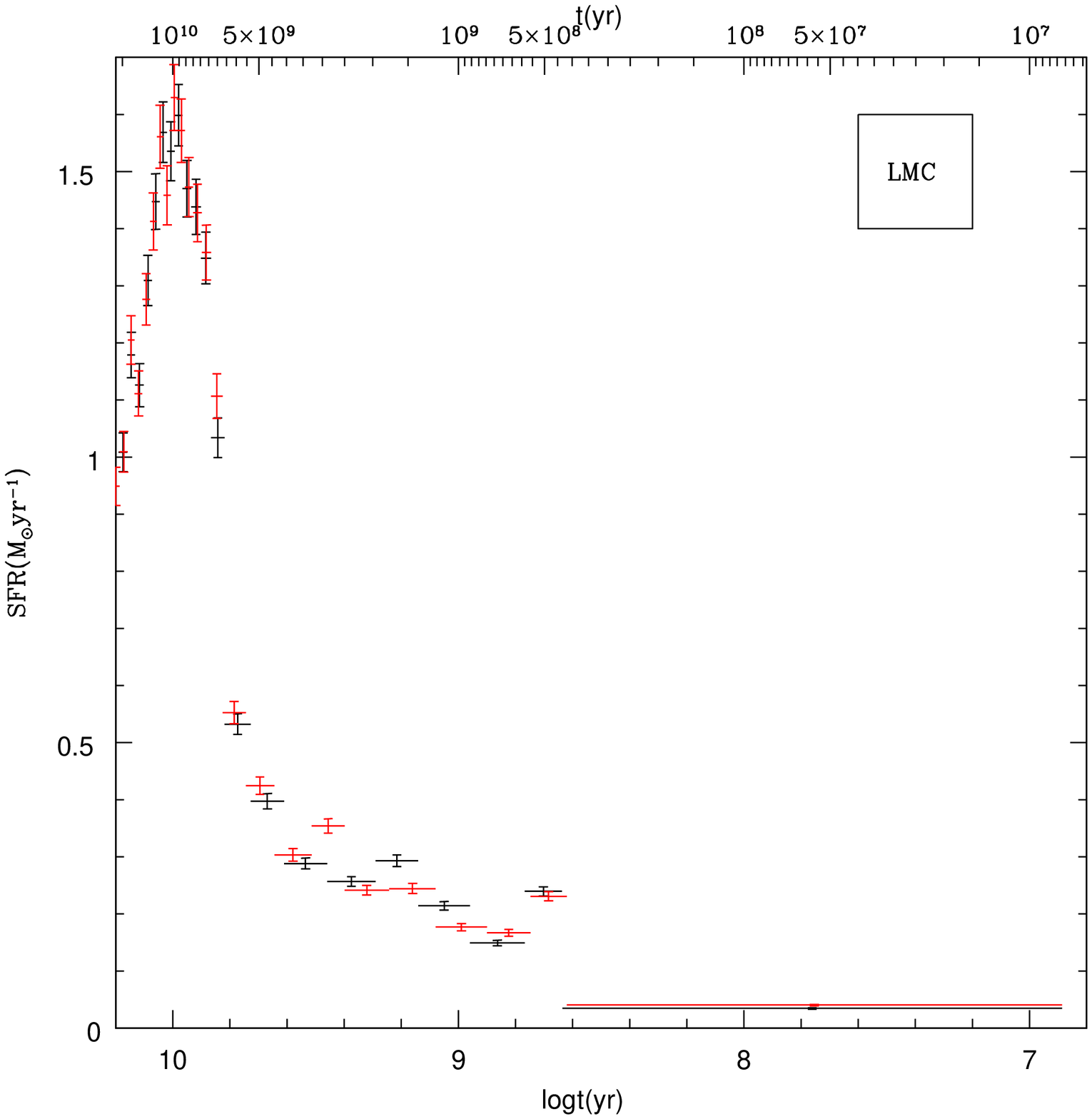,width=88mm}
\psfig{figure=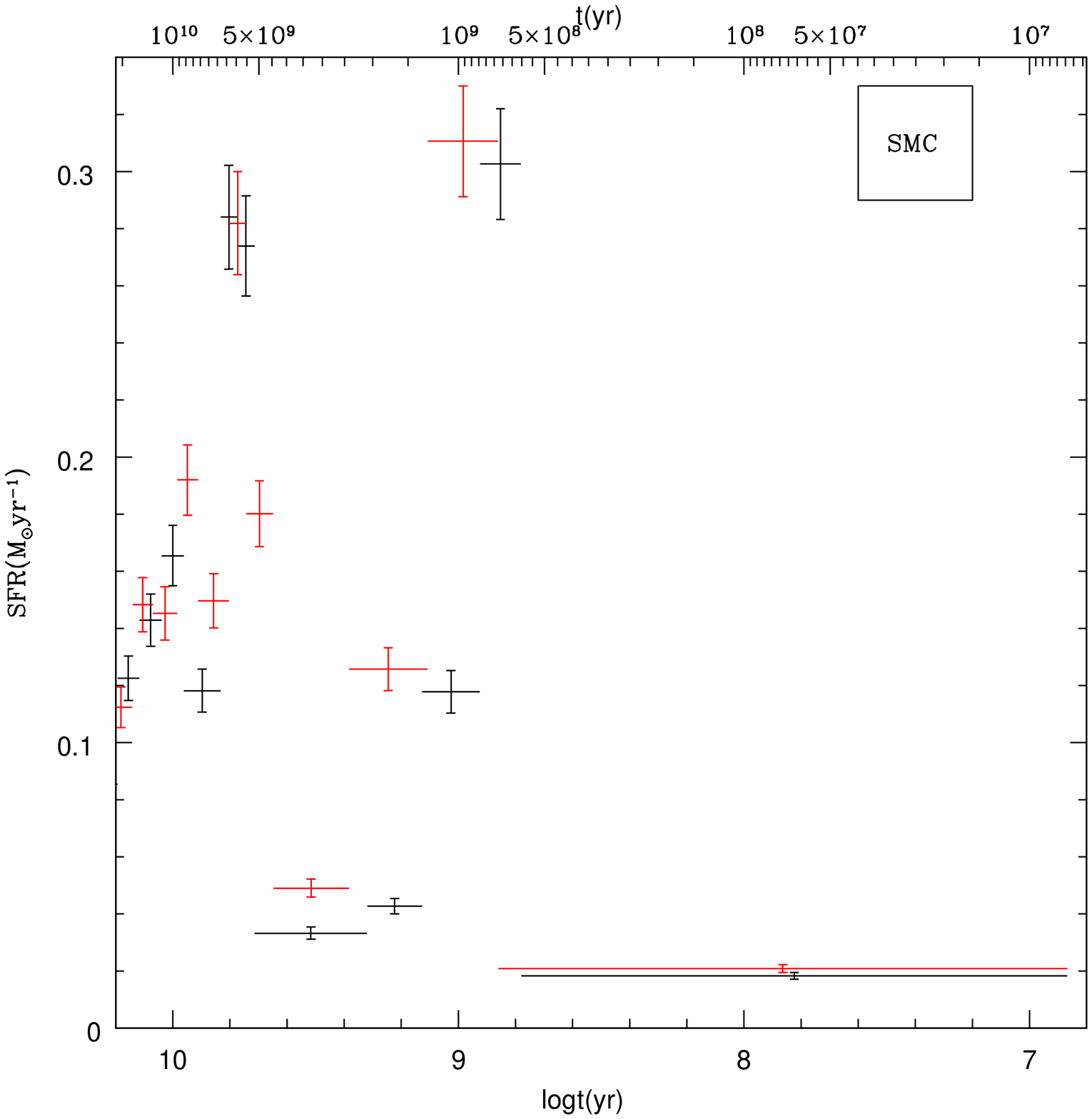,width=88mm}
}}
\caption[]{Global SFH of the LMC ({\it left}) and SMC ({\it right}) using both
reddening correction approaches; black lines represent applying a
carbonaceous-dust correction first, whilst red lines demonstrate the reverse
(see Section 3.1.1).  Vertical lines show statistical error bars described in Eq.\ (10). For the LMC, an initial burst of star formation as old as
10 Gyr is observed ($\log t=10$), with a SFR of $\sim1.5$ M$_\odot$ yr$^{-1}$,
followed by more quiescent star formation. In the case of the SMC, an initial
burst of star formation as old as 6 Gyr is observed ($\log t=9.8$), with a SFR
of $\sim0.28$ M$_\odot$ yr$^{-1}$, followed by a secondary star formation epoch
$\sim0.7$ Gyr ago ($\log t=8.8$) with a SFR of $\sim0.3$ M$_\odot$ yr$^{-1}$.}
\end{figure*}

We thus derive the SFRs as a function of time, where we group the stars into
bins of equal numbers -- i.e., equal Poissonian significance. Figure 4
represents the SFHs for both reddening correction approaches (see Section
3.1.1); black lines indicate the case in which a carbonaceous-dust correction
is first applied, and red lines demonstrate the reverse -- however, there is
no noticeable difference in the SFRs between these two correction approaches.  Horizontal lines show time bins while vertical lines demonstrate statistical error bars derived as following;
\begin{equation}
\sigma_\xi\ =\ \frac{\sqrt{N}}{N}\ \xi,
\end{equation}

where "N" is the number of stars in each age bin.
For the LMC (Fig.\ 4, left panel) we find an ancient star formation episode  with rate of $1.598\pm0.054$ M$_\odot$ yr$^{-1}$
$\sim10$ Gyr ago ($\log t=10$), corresponding to the formation of the LMC: the
total stellar mass produced in the LMC is  $\sim2.2\times10^{10}$ M$_\odot$, of
which $\approx89$ per cent was formed during the first epoch. For the SMC
(Fig.\ 4, right panel), two formation epochs are observed; one with a SFR of
$0.282\pm0.017$ M$_\odot$ yr$^{-1}$ rate $\sim6$ Gyr ago ($\log t=9.8$) and a
secondary star formation episode $\sim0.7$ Gyr ago ($\log t=8.8$) with a
similar SFR of $0.310\pm0.019$ M$_{\odot}$ yr$^{-1}$. These findings are in good
agreement with recent studies, e.g.\ Weisz et al.\ (2013). Of a total stellar
mass in the SMC of $\sim4.0\times10^8$ M$_\odot$, $\approx82$ per cent was
produced during the first peak in SFR, and $\approx11$ per cent during the
second peak.

\subsection{Star formation history of the central regions}

\begin{figure*}
\centerline{\hbox{
\psfig{figure=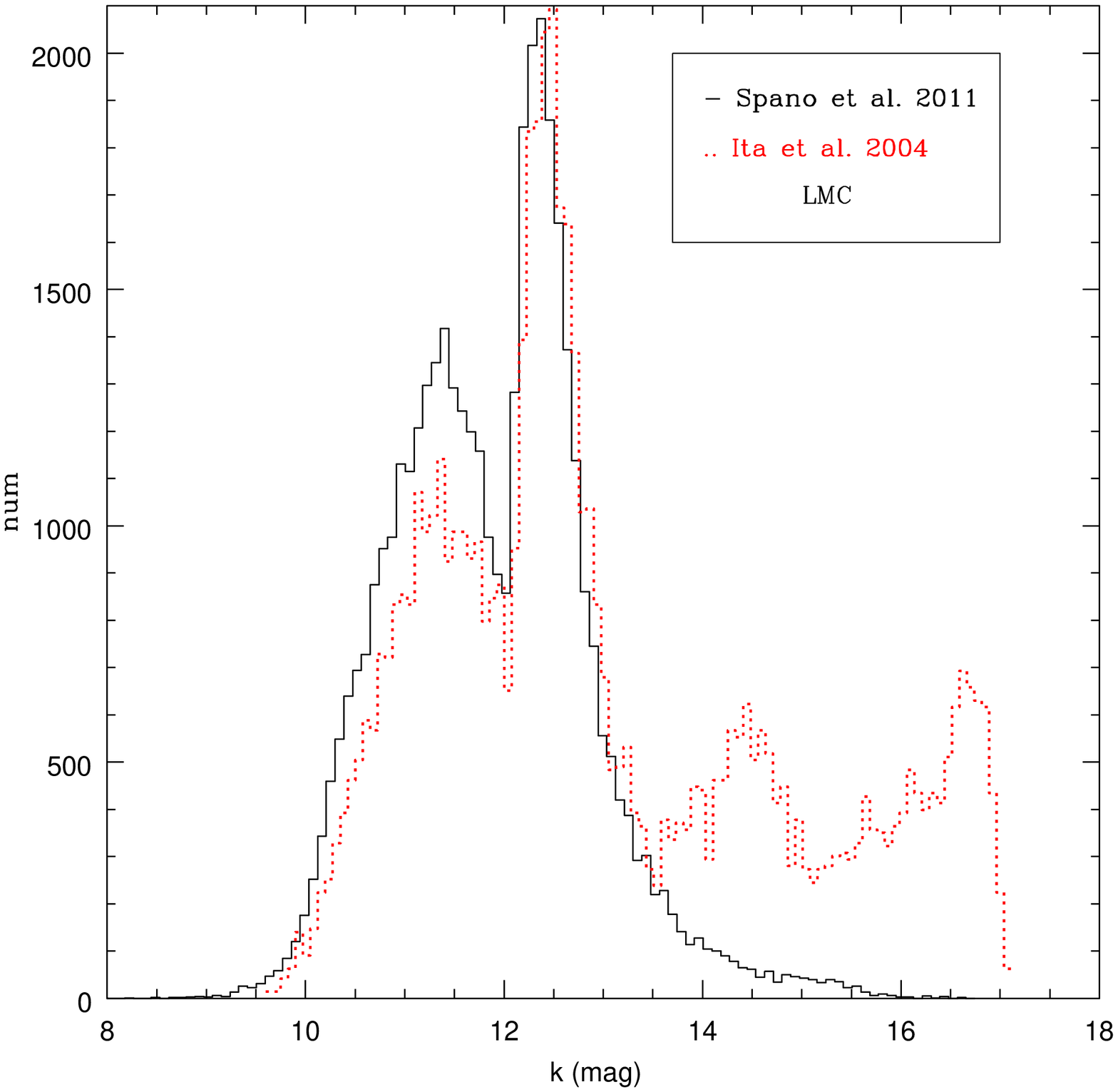,width=88mm}
\psfig{figure=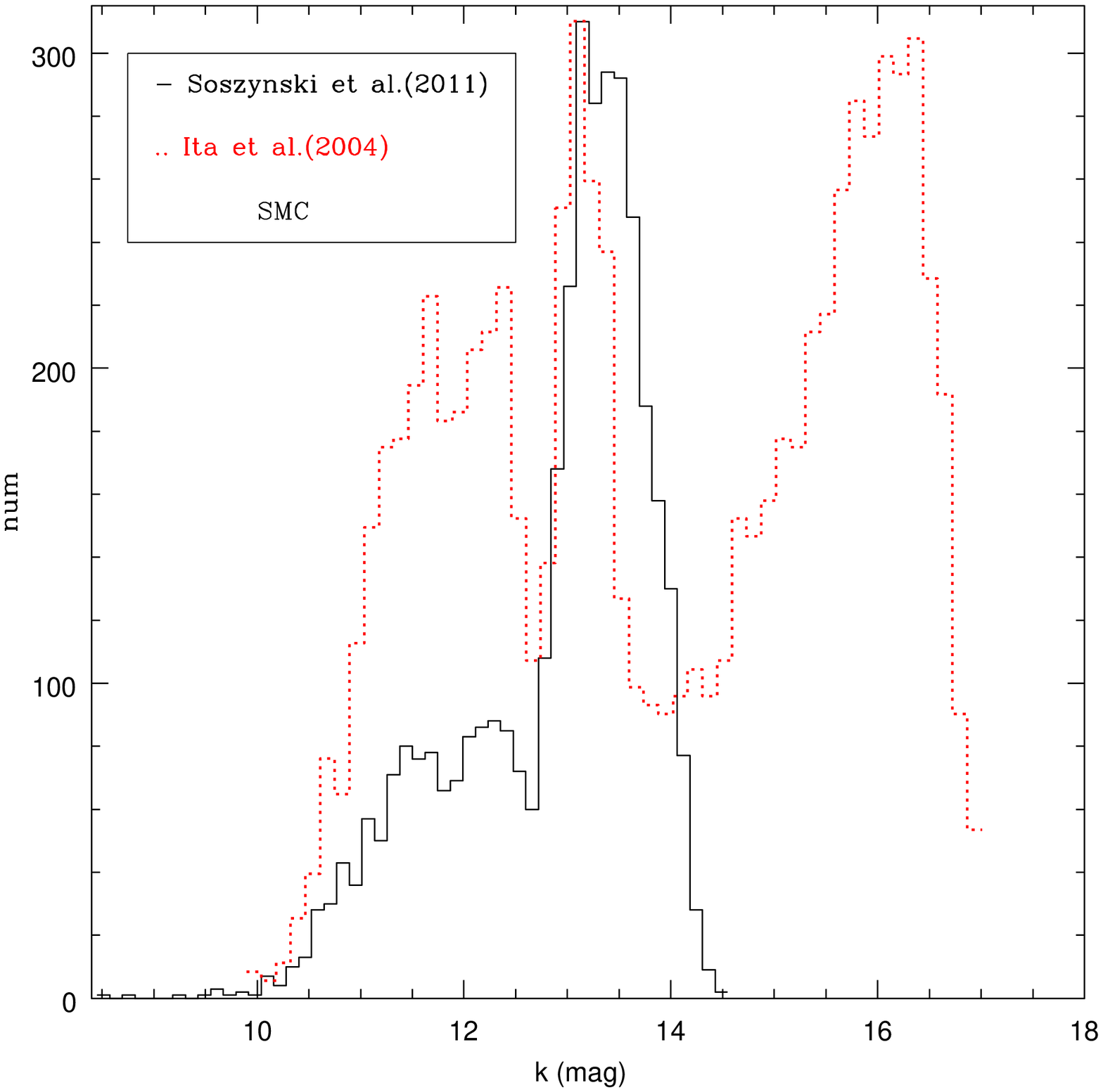,width=88mm}
}}
\caption[]{K-band histograms for variable stars in the overlapping regions
(see Fig.\ 1) in the central parts of the LMC ({\it left}) and SMC ({\it
right}), from Ita et al.\ (2004a) (red dotted lines), and from Spano et al.\
(2011) and Soszy\'nski et al.\ (2011) (solid black lines) in the LMC and SMC,
respectively. Each peak shows a type of variables, from left to right: AGB
stars, RGB stars and Cepheids.}
\end{figure*}

Before investigating the SFH in the central regions of the LMC and SMC, and
comparing it to the SFH in their peripheries, we consider the use of the Ita
et al.\ (2004a) catalogue for the central regions. Figure 5 shows the
normalised K-band luminosity distributions. The peak at the brightest K-band
magnitudes corresponds to the AGB; the second peak starts at $K\sim12$ mag for
the LMC and $K\sim12.7$ mag for the SMC, which corresponds to the RGB tip. The
deeper Ita et al.\ catalogue also contains fainter stars including Cepheids.
For the purpose of our study, based mainly on AGB stars (and small numbers of
RSGs) there is no indication that the IR catalogue of Ita et al.\ of the LMC
variables offers any major advantage over the Spano et al.\ (2011) catalogue
(we remind the reader that the latter is originally an optical catalogue cross
matched with 2MASS). On the other hand, in the central part of the SMC, the
Ita et al.\ catalogue is preferred over that of Soszy\'nski et al.\ (2011) as
the former is much more complete for the AGB stars.

\begin{figure*}
\centerline{\hbox{
\psfig{figure=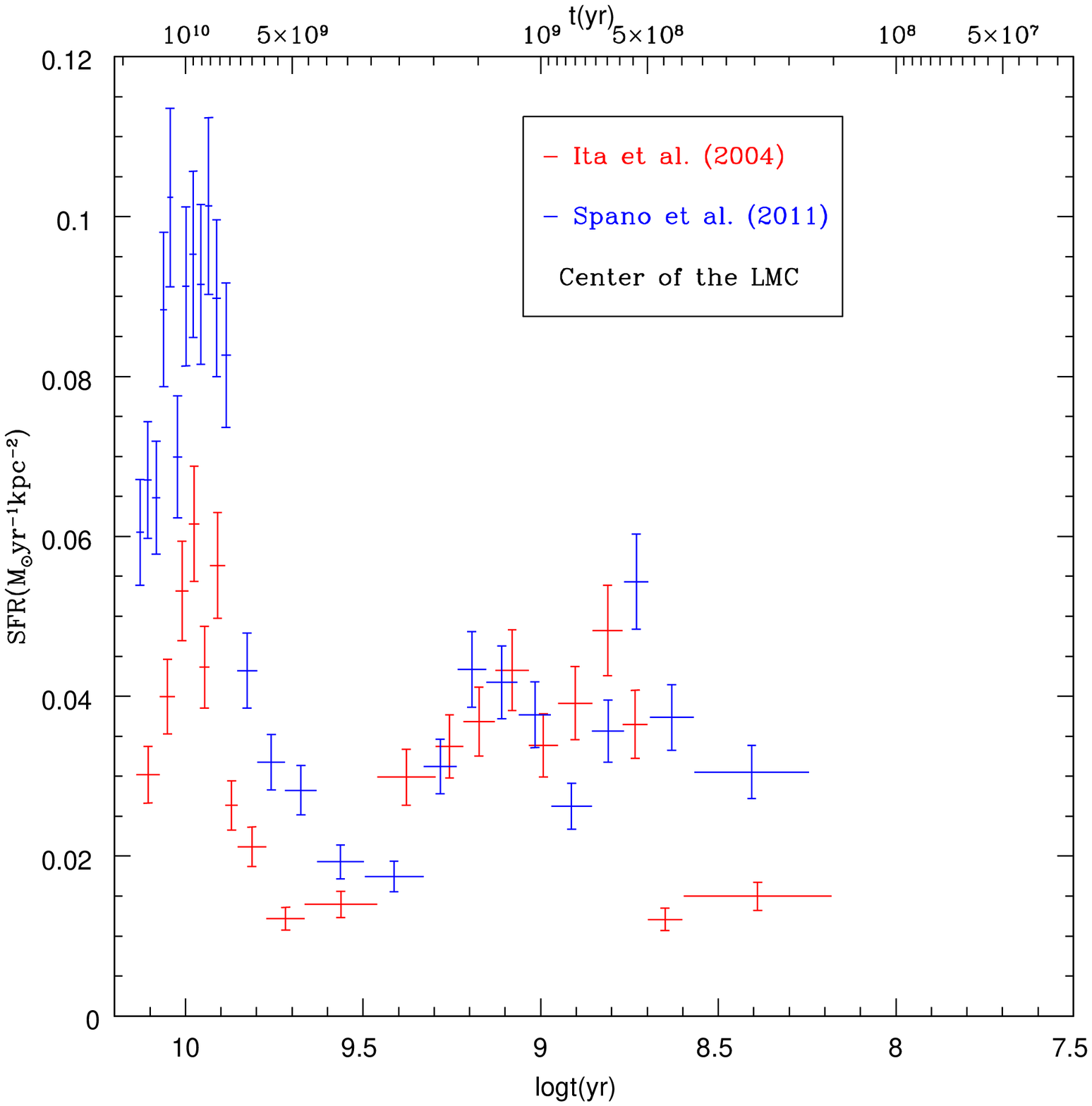,width=88mm}
\psfig{figure=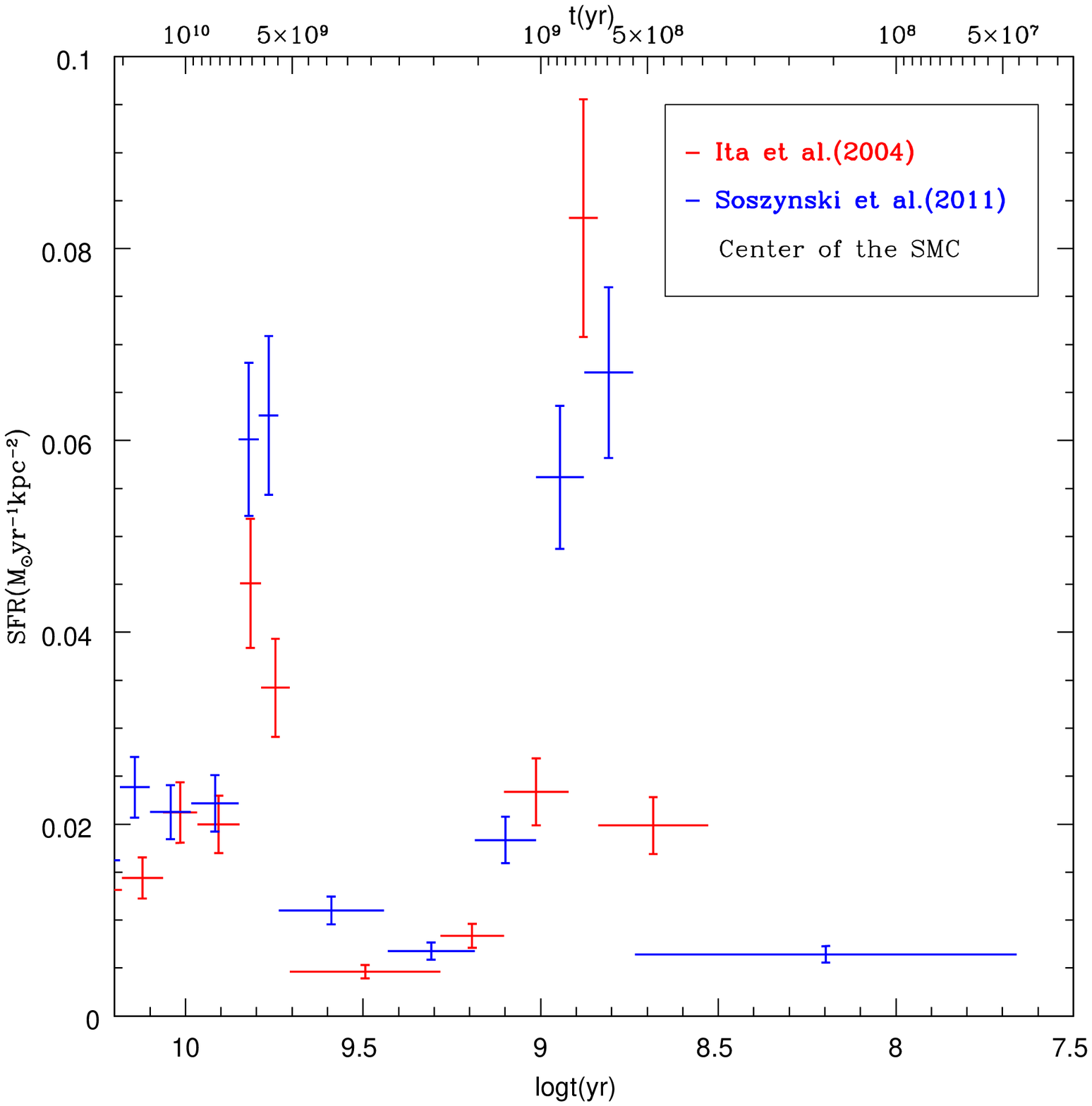,width=88mm}
}}
\caption[]{{\it Left:} SFH of the central bar structure in the LMC based on
the catalogues of Ita et al.\ (2004a; red symbols) and Spano et al.\ (2011;
blue symbols);  along with statistical error bars discussed in Section 4.1. The Spano et al.\ data indicate a higher rate of initial star
formation, by a factor of two compared to the Ita et al.\ data. However, the
two data sets agree well on the more recent SFRs. {\it Right:} SFH of the
overlapping region at the centre of the SMC based on the catalogues of Ita et
al.\ (2004a; red symbols) and Soszy\'nski et al.\ (2011; blue symbols). The
Ita et al.\ data imply a higher rate of secondary star formation, whilst the
Soszy\'nski et al.\ data imply relatively higher rates during the initial star
formation episode.}
\end{figure*}

The SFH of the LMC was derived from the Ita et al.\ (2004a) catalogue, and for
the overlapping region with the Spano catalogue (left panel in figure 6). This
area corresponds to the bar structure of the LMC (see Fig.\ 1). For a
meaningful comparison with the literature we report the SFR per physical area.
The SFR in the LMC bar varies from $\sim0.01$ to $0.1$ M$_\odot$ yr$^{-1}$
kpc$^{-2}$. Two formation epochs are seen; one epoch $\sim10$ Gyr ago ($\log
t=10$) when almost 48 per cent of the total stellar mass within the bar was
formed, and a more recent epoch starting $\sim3$ Gyr ago ($\log t=9.4$) which
lasted until as recent as 500 Myr ago ($\log t=8.7$), within which almost 30
per cent of the total stellar mass was formed (the total stellar mass in the
covered area is $\sim1.4\times10^8$ M$_\odot$). These two estimates are
consistent between the use of both catalogues, except for the ancient epoch in
which a higher SFR is found for the Spano data. However, it appears as though
the SFR based on Spano et al.\ (2011) is more successful in recovering the
ancient star formation.  Cioni et al. (2014) also found the star formation history of the LMC using Vista survey of the Magellanic Clouds (VMC) which reported an old star formation at around $\log t=9.9$ as well as a secondary peak of star formation at $\sim\log t=8.5-8.7$ for stars closer to the LMC center. These two peaks in SFR coincide with the peaks that we have determined, though we argue for a more sustained star formation over the most recent few Gyr.

The SFH of the SMC based on the Ita et al.\ (2004a) catalogue was also derived
and compared with that derived for the overlapping region in the Soszy\'nski
et al.\ (2011) catalogue (right panel in figure 6). This corresponds to the
central, approximately square-degree region of the SMC (see Fig.\ 1). The SFR
varies from $ \sim0.006$ M$_\odot$ yr$^{-1}$ kpc$^{-2}$ to $\sim0.07$ M$_\odot$
yr$^{-1}$ kpc$^{-2}$. As in other studies (e.g., Cignoni et al.\ 2012), two
formation episodes can be identified; one episode $\sim6$ Gyr ago ($\log
t=9.8$) when almost 49 per cent of the total stellar mass was formed, and
another one $\sim800$ Myr ago ($\log t=8.9$) within which around 12 per cent
of the total stellar mass was formed. As for the LMC, the SFRs obtained for
both catalogues agree rather well, however, the Ita et al.\ data indicate a
relatively more pronounced secondary (recent) episode of star formation.

\begin{figure*}
\centerline{\hbox{
\psfig{figure=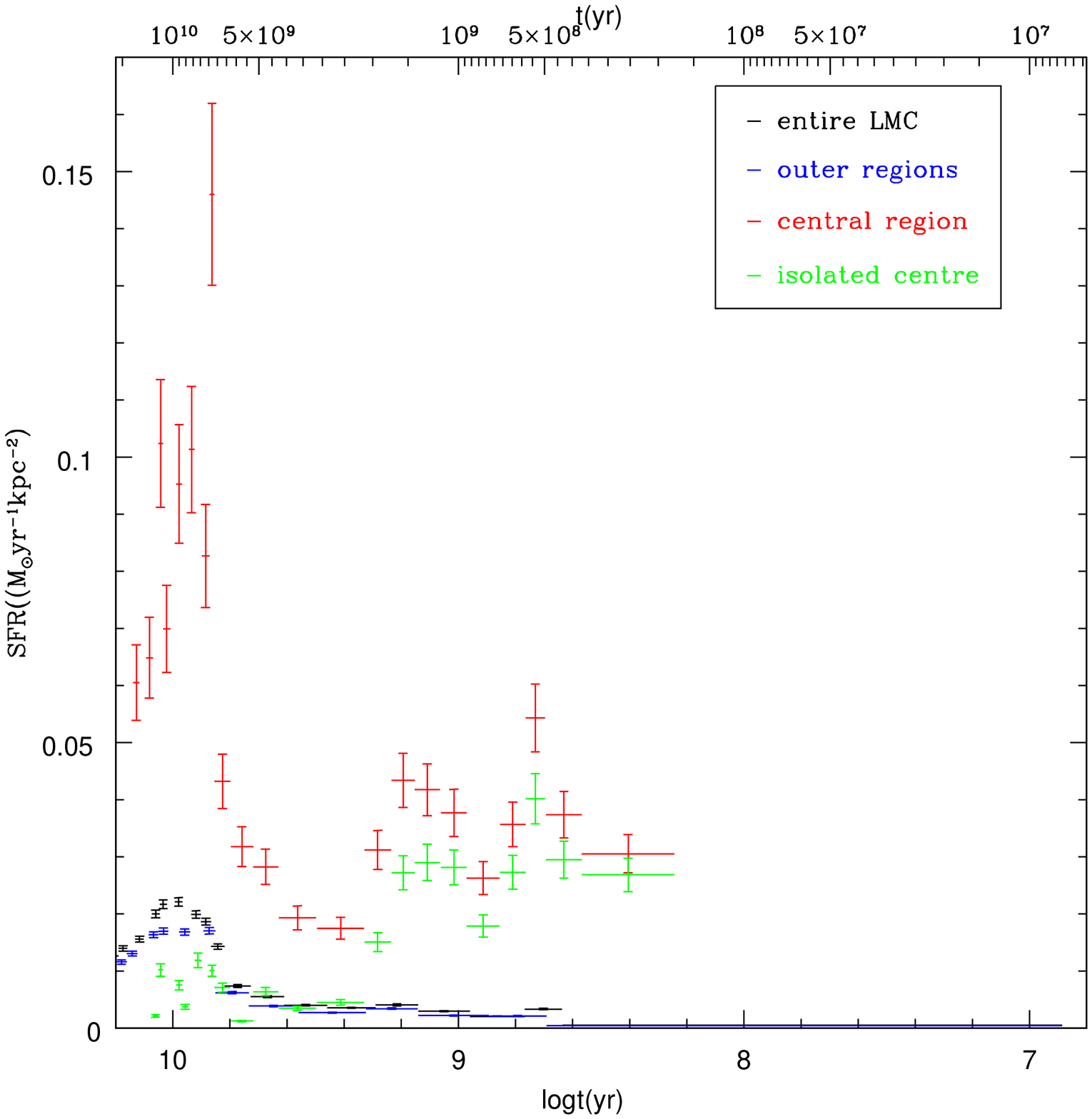,width=88mm}
\psfig{figure=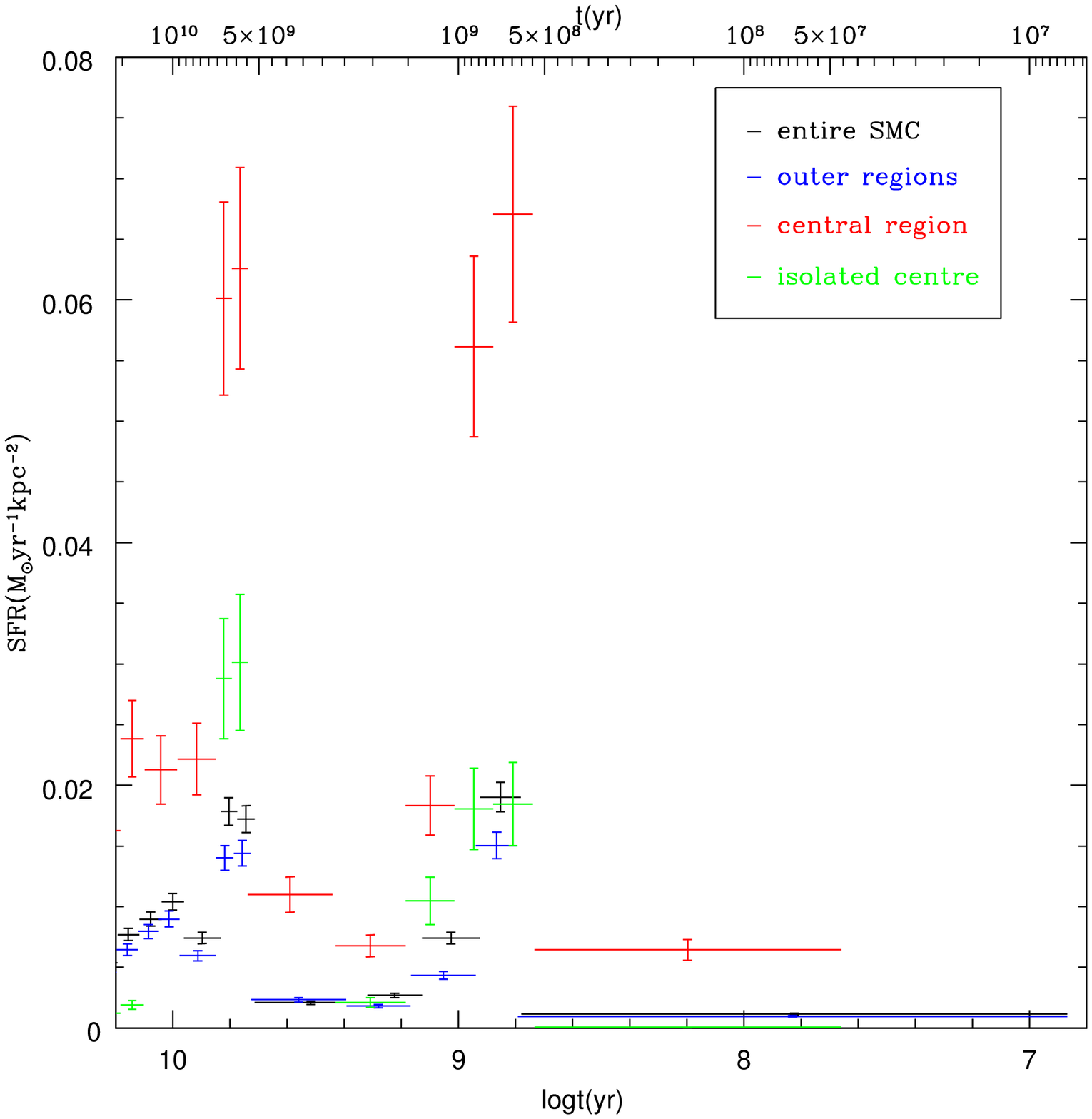,width=88mm}
}}
\caption[]{SFHs based on LPVs in the catalogues of Spano et al.\ (2011) for
the LMC ({\it left}) and Soszy\'nski et al.\ (2011) for the SMC ({\it right}).
Black symbols: global star formation; blue symbols: star formation in the
outskirt of the galaxies; red symbols: bar (for the LMC) and central (for the
SMC) star formation; and green symbols: isolated star formation for the
central regions derived by subtracting the SFH of the surrounding parts;  along with statistical error bars explained in Section 4.1. For
both galaxies, the bar/central star formation is remarkably dominant.}
\end{figure*}

By excluding the stars in the bar/central regions of the galaxies (amounting
to 3 deg$^{2}$ and 1 deg$^{2}$ in the LMC and SMC, respectively), we also
obtained the SFH for the outer regions. Conversely, we also isolated the SFH
of the bar/central regions by subtracting that derived for the immediately
surrounding regions (which could also be seen in projection against and/or
mixed with any distinct SFHs belonging to the bar/central regions). Figure 7
shows a comparison between the SFHs derived for the different regions.  In case of the LMC, the second peak of star formation is not seen in the outer region whilst for the SMC, another remarkable peak of star formation is still observed in outer region. The latter is in good agreement with the work by Ripepi et al.\ (2014) who, using the optical STEP survey, found regions of high levels of star formation in the North and in the Wing of the SMC.

\begin{figure}
\centerline{\psfig{figure=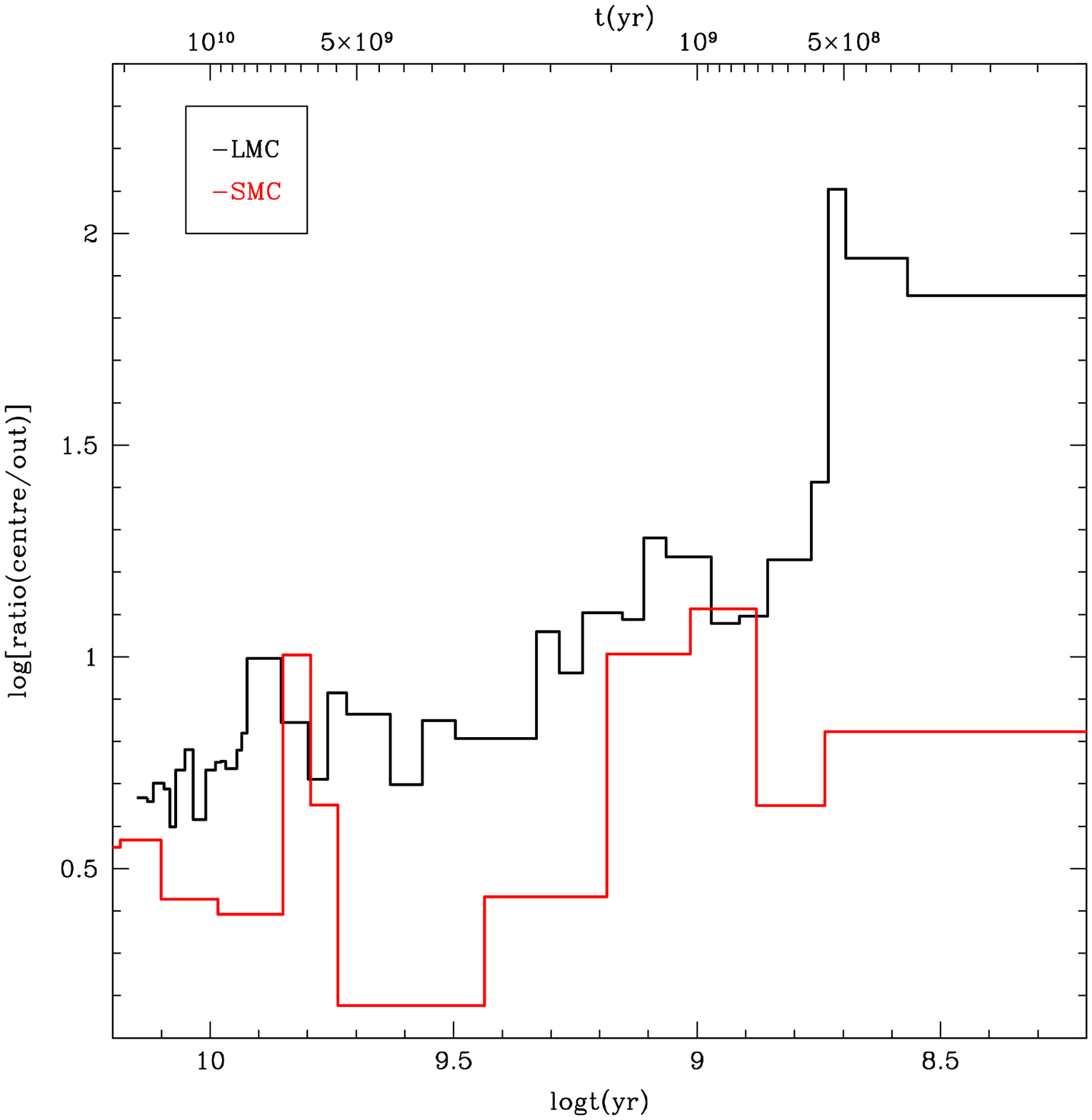,width=86mm}}
\caption[]{Ratio of the SFR in the centre compared to that in the outskirts,
for the LMC (in black) and SMC (in red). Each bin in the histogram contains
an equal number of stars. Clearly, star formation activity has been centrally
concentrated in recent times, and possible always if old stars have migrated
to the outskirts.}
\end{figure}

Figure 8 shows the ratio of the SFR in the central regions compared to that
in the outer regions. Each bin in the histograms contain an equal number of
stars. Note also the logarithmic scale. The LMC shows a steady concentration
of younger stars in the central regions.  A similar result was reported by Meschin et al.\ (2014), who found an ancient star formation epoch dating back 10 Gyr, followed by more recent star formation activity 4--1 Gyr ago. They also found younger stellar populations to be more centrally concentrated; in other words, the stellar population ages as one moves away from the center of the galaxy. Any such trend in the SMC is less
obvious, though might also be present. Dynamical effects will have resulted
in the migration of old stars towards the outskirts, so it is impossible to
tell on the basis of this work whether star formation activity has always
been centrally concentrated, or whether it has progressed further inwards as
the galaxies matured. As the dynamical timescale is not much more than 100
Myr, stars that were formed more than a Gyr ago will have equilibrated their
kinematics with the gravitational potential well -- unless they have been
exposed to recent tidal disturbances.

\section{Concluding remarks}

We find a significant difference in the ancient SFH of the LMC and the SMC.
For the LMC, the bulk of the stars formed $\approx10$ Gyr ago, while the
strongest episode of star formation in the SMC occurred a few Gyr later. This
is consistent with the findings of Weisz et al.\ (2013) obtained through CMD
analysis of optical (Hubble space Telescope) photometry obtained in a few
small regions in the centres and outskirts of the Magellanic Clouds. They
argued that the relative enhancement in the SFH of the SMC is a consequence
of the interaction with the LMC, with a stronger impact on the global star
formation of the SMC due to its much smaller mass compared to the LMC. The
smaller size of the SMC may also have resulted in the slower assembly at
early times, similar to what was found in M\,33 (Javadi et al.\ 2011). This
would be easier to understand if the SMC and LMC did not form as a pair, but
at least the SMC was formed in isolation.

A secondary peak of the SFH for the central part of the LMC has been reported
in other studies, e.g., Smecker-Hane et al.\ (2001) who found a dominant star
formation activity within the bar of the LMC at intermediate ages. Having
isolated the SFH of the bar of the LMC from that of the disc, it has become
clear that the bar formed a few Gyr ago and has maintained its star formation
until very recently ($\sim200$ Myr). However, the secondary peak in SFR in
the SMC is much sharper (in duration) and occurred $\sim700$ Myr ago. There
is indeed a similar peak noticeable in the SFH of the LMC, and this is thus
distinct from the star formation activity related to the bar itself. The
co-eval 700-Myr bursts in the SMC and LMC are arguably due to the tidal
interaction between the Magellanic Clouds and possibly their approach of the
Milky Way (cf.\ Bekki \& Chiba 2005; Indu \& Subramaniam 2011).

\begin{figure}
\centerline{\psfig{figure=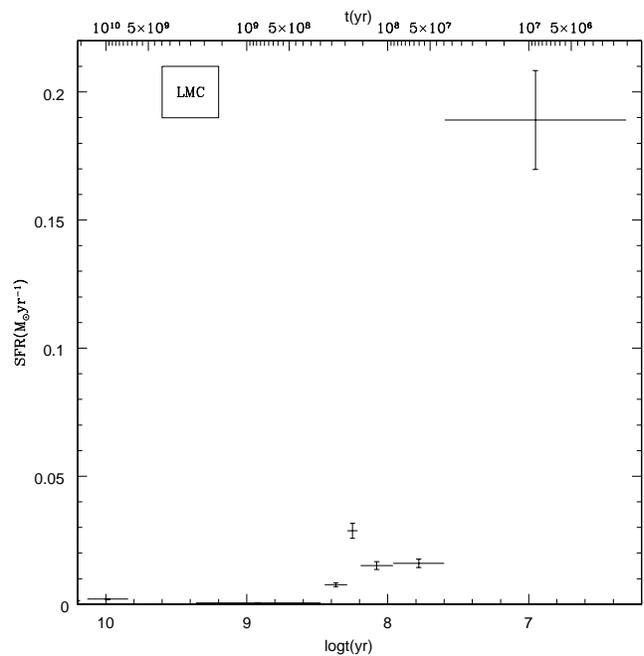,width=86mm}}
\caption[]{Recent star formation of the Large Magellanic Cloud using Ulaczyk et al.\ (2013) catalogue of more massive (brighter) stars.}
\end{figure}

The LMC catalogues we have worked with are limited to $t\sim100$ Myr due to
the lack of more massive stars (brighter stars) because of saturation of the
detectors used in the surveys. However, Ulaczyk et al.\ (2013) presented a
catalogue of brighter stars that had been saturated in the main OGLE
catalogue. On the basis of this bright sample we derive  the recent SFH in the LMC using the methods described above (Fig.\ 9). We find a SFR of
$\approx0.19$ M$_\odot$ yr$^{-1}$ at $t\sim5$ Myr, i.e.\ similar to that over
the past Gyr or so (compare, for instance, with figure 4).

After applying a correction to the pulsation duration of LPVs as predicted by
the stellar evolutionary models (see Javadi et al.\ 2013), the resulting SFRs
are now in very good agreement with those obtained in other studies (e.g.,
Harris \& Zaritsky 2004, 2009; Cignoni et al.\ 2012). However, a recent
investigation by the Padova group of their own models revealed a severe
over-estimation of the lifetimes of thermal-pulsing AGB stars (Girardi et
al.\ 2013) which could also have implications for the duration of the radial
pulsation phase. Indeed, the techniques we have pioneered in M\,33 (Javadi et
al.\ 2011, 2013), and now applied in the Magellanic Clouds, offer a promising
avenue towards exposing the assembly and development of galaxies as well as
improving models for the late stages of stellar evolution and mass loss.

\section*{Acknowledgments}
We wish to thank Spano et al., Soszy\'nski et al.\ and Ita et al.\ for making
their catalogues of variable stars public. This publication makes use of data
products from the Two Micron All Sky Survey, which is a joint project of the
University of Massachusetts and the Infrared Processing and Analysis
Center/California Institute of Technology, funded by the National Aeronautics
and Space Administration and the National Science Foundation.

\label{lastpage}
\end{document}